\documentclass{aa}  

\usepackage{graphicx}
\usepackage{txfonts}
\usepackage{lipsum}
\usepackage{subcaption}
\usepackage{lscape}    
                               
\usepackage{placeins}           

\usepackage{hyperref}

\begin{document}

   \title{Exploring the dynamical evolution of binary stars in multiple-population globular clusters}

   \author{J. Bruce\inst{1}\fnmsep\thanks{E-mail: jobruce@iu.edu}
        \and E. Vesperini\inst{1}
        \and A. Askar\inst{2}
        \and E. Bortolan\inst{4}
        \and M. Giersz\inst{2}
        \and J. Hong\inst{5}
        \and A. Hypki\inst{6}
        \and A. P. Milone\inst{3}\fnmsep\inst{4}
        }

   \institute{Department of Astronomy, Indiana University, Swain West, 727 E. 3rd Street, Bloomington, IN 47405, USA
            \and Nicolaus Copernicus Astronomical Center, Polish Academy of Sciences, ul. Bartycka 18, 00-716 Warsaw, Poland
            \and Istituto Nazionale di Astrofisica - Osservatorio Astronomico di Padova, Vicolo dell'Osservatorio 5, Padova 35122, Italy
            \and Dipartimento di Fisica e Astronomia "Galileo Galilei", Univ. di Padova, Vicolo dell'Osservatorio 3, Padova 35122, Italy 
            \and Korea Astronomy and Space Science Institute, Daejeon 34055, Republic of Korea
            \and Faculty of Mathematics and Computer Science, A. Mickiewicz University, Uniwersytetu Pozna\'nskiego 4, 61-614 Pozna\'n, Poland\\ }

   \date{Accepted January 25, 2026}
 
  \abstract
    {\noindent The presence of multiple stellar populations in globular clusters leads to a complex dynamical environment that significantly influences the evolution of binary stars, which in turn impacts the evolution of the cluster itself. For this study, we used a series of Monte Carlo simulations run with the MOCCA code to investigate the long-term dynamical evolution of binary stars in globular clusters hosting two distinct stellar populations. We explored how global binary properties such as incidence, fraction, and spatial distribution evolve over time due to the unique dynamical environment associated with each population. Our results show how binaries in the more centrally concentrated second population (P2) experience increased rates of hardening and disruption relative to the first population (P1), leading to distinct radial profiles in binary incidence and fraction. We also demonstrate the difference in spatial mixing timescales for binaries compared to single stars, where binary stars in each population retain some memory of their initial configurations even after complete single star mixing. Additionally, we investigated the formation and evolution of mixed binaries (binaries composed of a P1 component and a P2 component), which form primarily within the core through dynamical interactions. Finally, we studied main sequence--white dwarf binaries and find that they represent a larger fraction of binaries in P1 compared to P2. The results of this paper highlight the interplay between cluster dynamics and the evolution of binary stars and how binaries can act as tracers of the cluster's initial conditions and dynamical evolution.}

   \maketitle
   \nolinenumbers

\section{Introduction}

Galactic globular clusters (GCs) were traditionally considered to be simple stellar populations. However, spectroscopic and photometric studies have shown that the majority of GCs host at least two stellar populations, characterized by variations in their light element abundances (see, e.g., \citeauthor{Milone+2022} \citeyear{Milone+2022} for a recent review). The origin and dynamical evolution of these multiple populations are not yet completely understood, and is an important open question in the field. 

Several theoretical models have been proposed regarding the formation of multiple populations, most of which involve the second population (P2) forming from a combination of reaccreted pristine gas and a pollutant such as ejecta from asymptotic giant branch (AGB) stars \citep{Ventura+2001, Calura+2019}, supermassive stars \citep{Gieles+2018}, or massive binary stars \citep{deMink+2009, Bastian+2013, Nguyen+Sills2024}. A common prediction of these models is that P2 stars form in a more concentrated central subsystem within a more extended P1; this has been observed in numerous clusters (see, e.g., \citeauthor{Bellini+2009} \citeyear{Bellini+2009}; \citeauthor{Lardo+2011} \citeyear{Lardo+2011}; \citeauthor{Beccari+2013} \citeyear{Beccari+2013}; \citeauthor{Cordero+2014} \citeyear{Cordero+2014}; \citeauthor{simioni+2016} \citeyear{simioni+2016}; \citeauthor{Boberg+2016} \citeyear{Boberg+2016}; \citeauthor{Dalessandro+2019} \citeyear{Dalessandro+2019}; \citeauthor{Onorato+2023} \citeyear{Onorato+2023}; \citeauthor{Cadelano+2024} \citeyear{Cadelano+2024}).

The initial structural differences between the multiple populations has many important implications for the dynamical evolution of the cluster, particularly in the case of binary stars. The increased stellar density of P2 results in shorter local relaxation timescales and increased encounter rates, affecting the properties and overall survival of binary systems. Previous studies have shown that these environmental differences can have significant effects on the survival of binaries and the evolution of their dynamical properties \citep{Vesperini+2011, Hong+2015, Lucatello+2015, Hong+2016, Dalessandro+2018, Hong+2019, kamann+2020, Milone+2020, Hypki+2022, Sollima+2022, Milone+2025, Bortolan+2025}.

For this paper we investigated the long-term dynamical evolution of binary stars in multiple-population globular clusters using a series of Monte Carlo simulations run with the MOCCA code \citep{Giersz1998, Hypki+2013, Hypki+2022, Hypki+2025}. We focused on how the distinct environments inhabited by binaries in each population affect the local and global binary fractions and incidences, the differing degrees of spatial mixing relative to the single stars of each population, and the formation and evolution of more exotic binaries. This paper builds on and extends previous studies by focusing on the evolution of the P1 and P2 binary incidences, the formation and dynamics of mixed binaries, and main sequence--white dwarf binaries (MS-WD) and their radial variation within a cluster. 

The structure of the paper is as follows. Section \ref{sec:2} details the simulation setup for this paper. In Section \ref{sec:3} we report the results obtained through analyzing the binary fractions and incidences (Section \ref{sec:3.1}), spatial mixing of the populations (\ref{sec:3.2}), as well as the formation and evolution of mixed binaries (\ref{sec:3.3}) and main sequence-white dwarf binaries (\ref{sec:3.4}). Finally, Section \ref{sec:4} summarizes and discusses the main findings of the paper.

\begin{table*}[ht]
\centering
\caption{Initial simulation parameters.}
\begin{tabular}{lcccc}
\hline
\hline
Simulation & $F^{P2}$ & Concentration Parameter & Binary Fraction & Initial Tidal Radius (pc) \\
\hline
mr025c01fb5 & 25\% & 0.1 & 5\% & 59.1 \\
mr025c01fb10 & 25\% & 0.1 & 10\% & 59.1 \\
mr025c005fb5 & 25\% & 0.05 & 5\% & 59.1 \\
mr025c005fb10 & 25\% & 0.05 & 10\% & 59.1 \\
mr01c01fb10 & 10\% & 0.1 & 10\% & 59.1 \\
mr025c01fb10sf & 25\% & 0.1 & 10\% & 38.0 \\
mr025c005fb10sf & 25\% & 0.05 & 10\% & 38.0 \\
\hline
\end{tabular}
\tablefoot{The metallicity (Z) of all simulations is initially set to 0.001, and we define the concentration parameter as the ratio of the initial half-mass radii for the two populations (${r_{\rm h}^{\rm P2}}/{r_{\rm h}^{\rm P1}}$). $F^{P2}$ is the fraction of the total number of objects (single or binary stars) in the P2 population.}
\label{tab:tab 1}
\end{table*}

\begin{table}[ht]
\centering
\caption{Simulation parameters at 12 Gyr.}
\begin{tabular}{lcccc}
\hline
\hline
Simulation & $M_{P2}/M_{Tot}$ & $R_{hl}$ (pc) & N\\
\hline
mr025c01fb5 & 0.567 & 3.026 & $3.519\times10^5$ \\
mr025c01fb10 & 0.574 & 2.763 & $3.313\times10^5$ \\
mr025c005fb5 & 0.601 & 2.737 & $3.163\times10^5$ \\
mr025c005fb10 & 0.618 & 2.527 & $2.936\times10^5$ \\
mr01c01fb10 & 0.444 & 3.830 & $1.561\times10^5$ \\
mr025c01fb10sf & 0.601 & 1.493 & $2.382\times10^5$ \\
mr025c005fb10sf & 0.655 & 1.158 & $1.856\times10^5$ \\
\hline
\end{tabular}
\tablefoot{We report the values of the clusters' half-light radii ($R_{hl}$) in pc, the number of stars remaining in the cluster (N), and the fraction of the total mass of the clusters in P2 stars ($M_{P2}/M_{Tot}$).}
\label{tab:tab 2}
\end{table}

\section{Methods}
\label{sec:2}

This study analyzes simulations carried out using the Monte Carlo Cluster Simulator (MOCCA) \citep{Giersz1998, Hypki+2013} code, a Monte Carlo-based tool designed to efficiently model the long-term dynamical evolution of globular clusters. A more detailed description of the code is provided in \citet{Hypki+2022, Hypki+2025}. Briefly, MOCCA incorporates stellar and binary evolution through the SSE and BSE algorithms (\citeauthor{Hurley+2000} \citeyear{Hurley+2000}, \citeyear{Hurley+2002}, with further updates from \citeauthor{kamlah+2022} \citeyear{kamlah+2022}) and models two-body relaxation through a Monte Carlo approach. Strong dynamical interactions are performed using the FEWBODY code \citep{Fregeau+2004, Fregeau+2007}. The choice of initial conditions for the simulations was motivated by hydrodynamical simulations (see, e.g., \citeauthor{Dercole+2008} \citeyear{Dercole+2008}; \citeauthor{Bekki2010} \citeyear{Bekki2010}, \citeyear{Bekki2011}; \citeauthor{Calura+2019} \citeyear{Calura+2019}) of multiple-population formation, in which P2 stars form in a centrally concentrated subsystem enveloped by a more extended P1 distribution.

The simulation setup follows initial conditions used in previous studies of multiple-population dynamics with both populations included in the system at the beginning of the simulation \citep{Hong+2015, Hong+2016, Vesperini+2021, Hypki+2022}. Each simulation contains a total of one million stars, with a binary fraction of 5\% or 10\% and a fraction of the total number of objects (where an object is defined as a single or binary star) in the P2 population equal to 10\% or 25\%. The initial metallicity is set to $Z=0.001$ in all simulations. The degree of central concentration for P2 is controlled by the concentration parameter, defined as the ratio of the P2 to P1 half-mass radii, which is set to be either 0.1 or 0.05. All models have an initial ratio of the half-mass radius to the tidal radius ($r_{\rm h}/r_{\rm t}$) equal to about 0.14 (except for the mr01c01fb10 model for which this ratio is equal to about 0.17). The cluster is initially in virial equilibrium with each population initially modeled with a King distribution \citep{King1966}, using central concentration parameters ($W_{\rm 0}$) of 5 or 7 for P1 and P2 respectively.

We explored binary fractions of either 5\% or 10\%, with the same fraction initially set for both populations. Stellar masses were drawn from a \citet{Kroupa2001} IMF between 0.1 and 100 $M_\odot$. Binary star pairing followed a uniform mass ratio distribution (0.1 $<$ q $<$ 1) for systems with primary star masses above 5 $M_\odot$, and random pairing otherwise based on the results of studies such as \citet{Kiminki+2012, Sana+2012, Kobulnicky+2014}. Semimajor axes for binaries are drawn from a uniform distribution in log(a), ranging from a minimum value of 

\begin{equation}
    a = 2\;\frac{(R_1 + R_2)}{1-e}
\end{equation}

\noindent where $R_1$ and $R_2$ are the stellar radii of the two stars and $e$ is the binary eccentricity, and 100 au. 

Clusters have an initial tidal radius equal to 59.1 pc (corresponding, for example, to systems moving on circular orbits around a point-mass galactic potential at a distance of 3.3 kpc). Two simulations have also been run for more compact systems with an initial tidal radius equal to 38.0 pc (corresponding to systems on a circular orbit at a galactocentric distance of 1.7 kpc); the shorter relaxation time of these systems along with the stronger tidal field causing them to lose more stars during their evolution implies that these systems reach an older dynamical age at the end of the simulations.

A summary of the initial conditions for all simulations is provided in Table \ref{tab:tab 1}. In Table \ref{tab:tab 2} we report the final values (at $t=12$ Gyr) of the clusters' number of stars, half-light radii, and fraction of the total cluster mass in P2 stars. We note that our suite of simulations is not designed to span the whole range of observed properties, but the final values of our models are consistent with those observed in many Galactic clusters.

\section{Results}
\label{sec:3}
\subsection{Binary fraction and incidence}
\label{sec:3.1}

\begin{figure}
    \centering
    \includegraphics[width=0.95\linewidth]{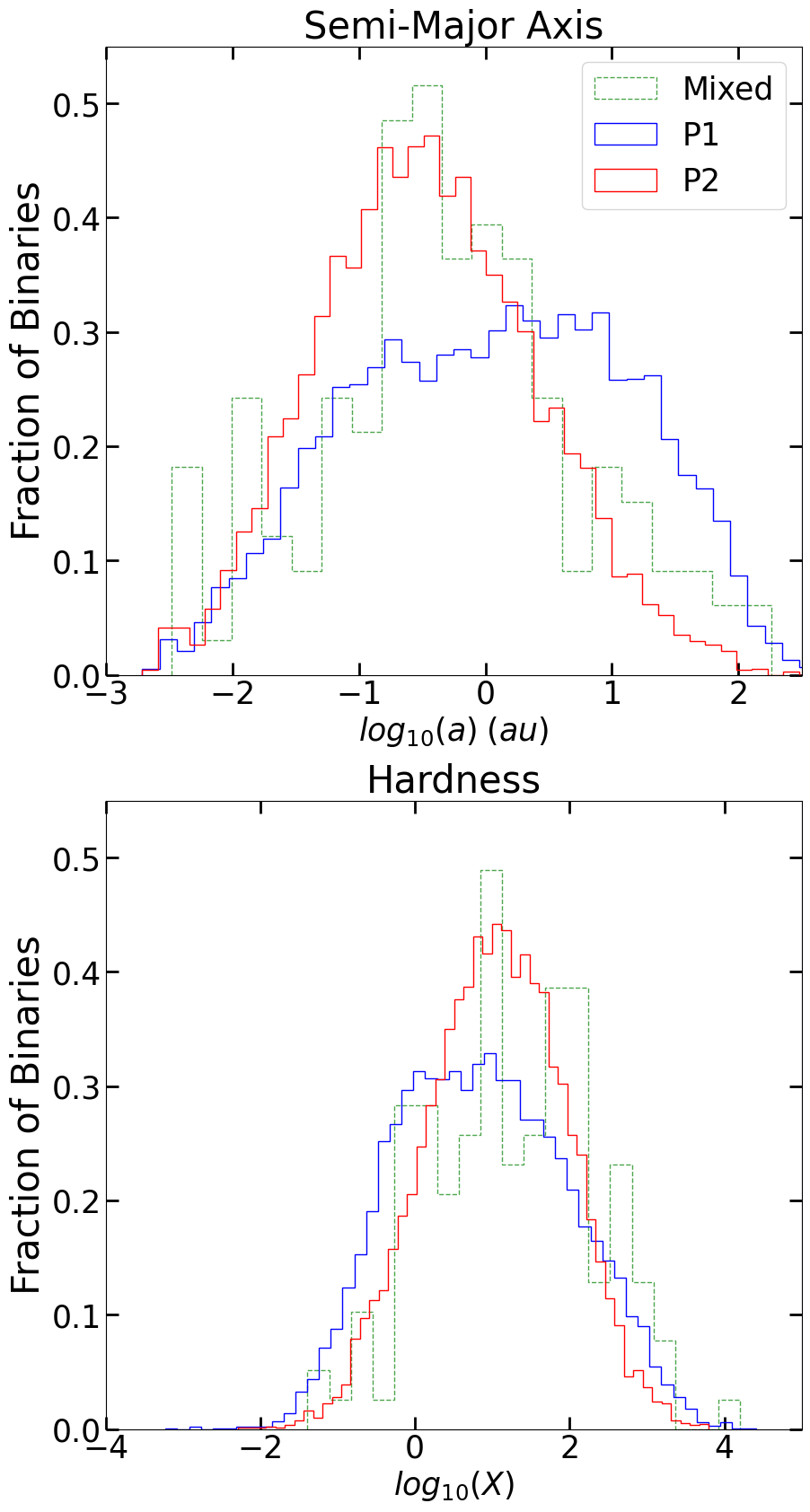}
    \caption{Distributions of the semimajor axes (top) and hardness values (bottom) of the P1 (blue), P2 (red), and mixed (green) binaries for the mr025c005fb10 simulation evolved to 12 Gyr and normalized such that the total area under each histogram equals 1.}
    \label{fig:fig 1}
\end{figure}

Figure \ref{fig:fig 1} shows the semimajor axis and hardness distributions of binaries at 12 Gyr for both P1 and P2 in simulation mr025c005fb10. The hardness of a binary is defined as 

\begin{equation}
    x = \frac{E_b}{m\sigma^2}
\end{equation}

\noindent where $E_b$ is the absolute value of the binary binding energy, $m$ is the average stellar mass, and $\sigma$ is the local one-dimensional velocity dispersion. Following Heggie's law \citep{Heggie1975}, binaries with $x>1$ are considered hard and tend to become harder through interactions, while soft binaries ($x<1$) become softer and are preferentially disrupted.

\begin{figure}
    \centering
    \includegraphics[width=0.95\linewidth]{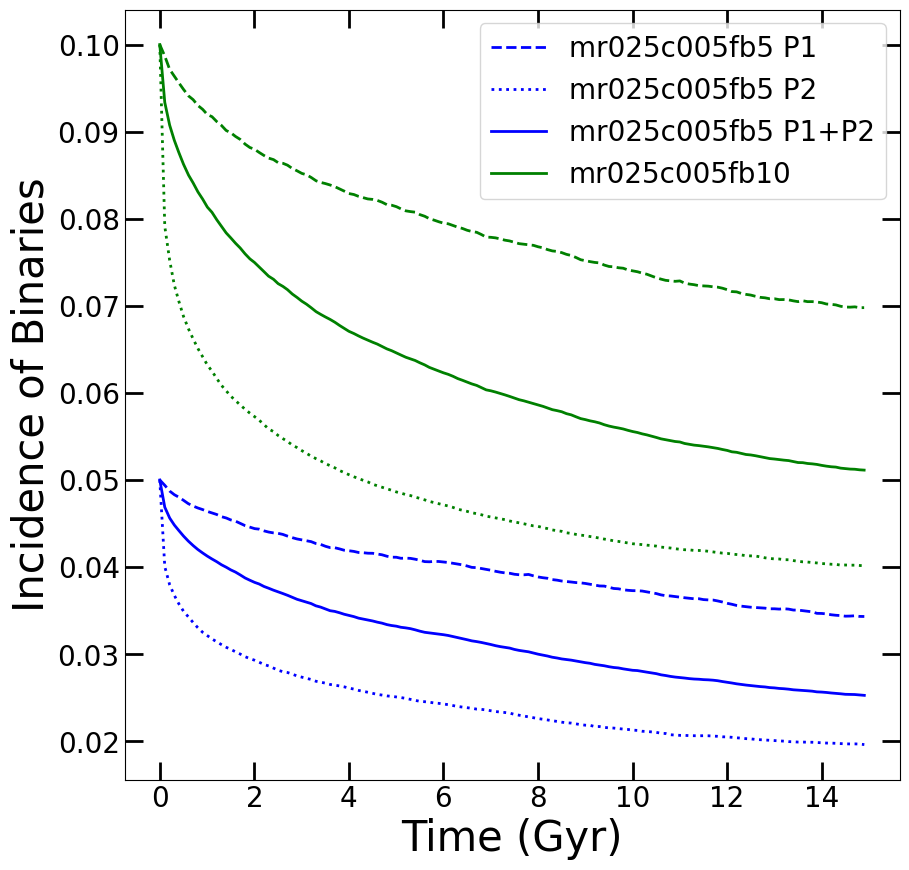}
    \caption{Time evolution of the global binary incidence within P1 (dashed line), within P2 (dotted line), and within all the stars in the cluster (P1+P2) (solid line). Incidence refers to the fraction of objects within each population that are binaries (see Eq. 3).}
    \label{fig:fig 2}
\end{figure}

P1 and P2 binaries have the same initial semimajor axis distribution. However, the dynamical evolution of the cluster leads to a divergence of the distributions as shown in Figure \ref{fig:fig 1}. In the less concentrated P1 environment, wider binaries survive longer due to lower encounter rates and larger local hardness (i.e. smaller local velocity dispersion at a given radius). Consequently, the interaction strength is insufficient to disrupt the binaries. In contrast, the centrally concentrated P2 experiences more frequent dynamical encounters, rapidly disrupting the wider binaries. As a result, P1 retains a larger population of wider binaries, while P2 becomes increasingly dominated by more compact binaries.

The divergence is similarly reflected in the hardness distributions. P2 contains a greater fraction of hard binaries, while P1 shows a larger proportion of soft binaries. These trends are consistent with the differences in the binary dynamics between the inner dense regions populated mainly by P2 stars, and the outer low-density regions populated mainly by P1 stars. In the inner regions, soft binaries are quickly disrupted and hard binaries are further hardened at an enhanced rate. Alternatively, in the P1-dominated outer regions both hardening and disruption of binaries proceed at a slower rate; moreover, the lower local velocity dispersion of the cluster outer regions implies that many binaries that would be soft binaries in the inner regions have larger hardness parameters and fall in the hard-binary regime.

The global impact on binary survival is shown in Figure \ref{fig:fig 2}, which displays the time evolution of the binary incidences within each population. Here we define the binary incidence for P1 ($I_b^{P1}$)

\begin{equation}
    I_b^{P1} = \frac{N_b^{P1}}{N_b^{P1}+N_s^{P1}}
\end{equation}

\noindent as the fraction of objects within P1 ($N_s^{P1}\;+\;N_b^{P1}$) that are binaries ($N_b^{P1}$). An analogous definition holds for P2. We note that the figures displayed in this paper show the results of all stars and binaries, regardless of stellar type, unless otherwise specified. However, when restricting the sample to main-sequence stars and binaries with $q > 0.5$ (as is often the case in observational studies; see, e.g., \citeauthor{Milone+2012} \citeyear{Milone+2012}), we recover similar qualitative trends across all figures. Initially, the binary incidence is identical in both populations. As the cluster evolves, P2 experiences a sharper decline in incidence due to the mechanisms described above, whereas P1 shows a shallower decrease. 

\begin{figure}
    \centering
    \includegraphics[width=0.95\linewidth]{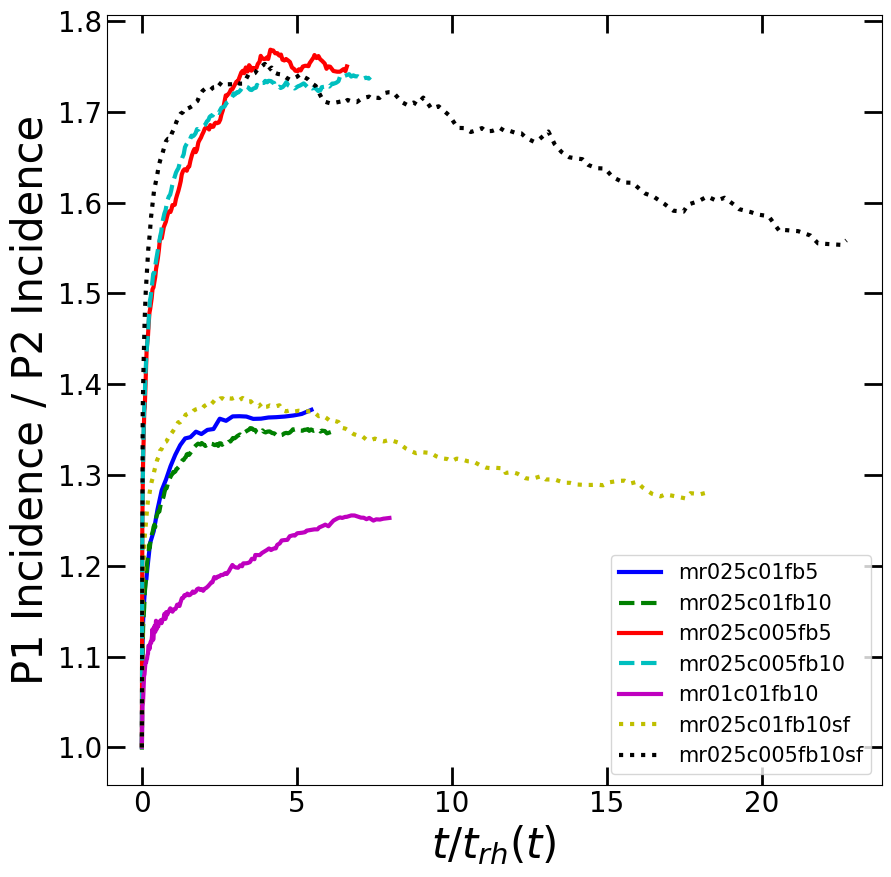}
    \caption{Time evolution of the ratio of the global P1 binary incidence to the global P2 binary incidence, normalized to the current half-mass relaxation time ($t_{rh}(t)$). Incidence refers to the fraction of objects within each population that are binaries (see Eq. 3).}
    \label{fig:fig 3}
\end{figure}

This difference in binary incidence evolution becomes clear in Figure \ref{fig:fig 3}, when we consider the time evolution of the ratio of the global P1 binary incidence to the global P2 binary incidence. The sharper decline in P2 incidence leads to an increase in the ratio for all simulations. This effect is emphasized in the simulations where P2 is initially even more concentrated. As the stellar populations mix and the binaries from each population are distributed within similar environments, the ratio flattens and stabilizes. For the two systems evolving in a stronger tidal field and reaching older dynamical ages (mr025c01fb10sf and mr025c005fb10sf) the ratio slightly declines during the late advanced phases of the cluster's evolution. At this stage, the remaining P2 binaries are very hard and thus remain largely unchanged in their incidence, whereas the soft P1 binaries in the outer regions begin to segregate toward the center and are rapidly disrupted by dynamical interactions.

\begin{figure}
    \centering
    \includegraphics[width=0.95\linewidth]{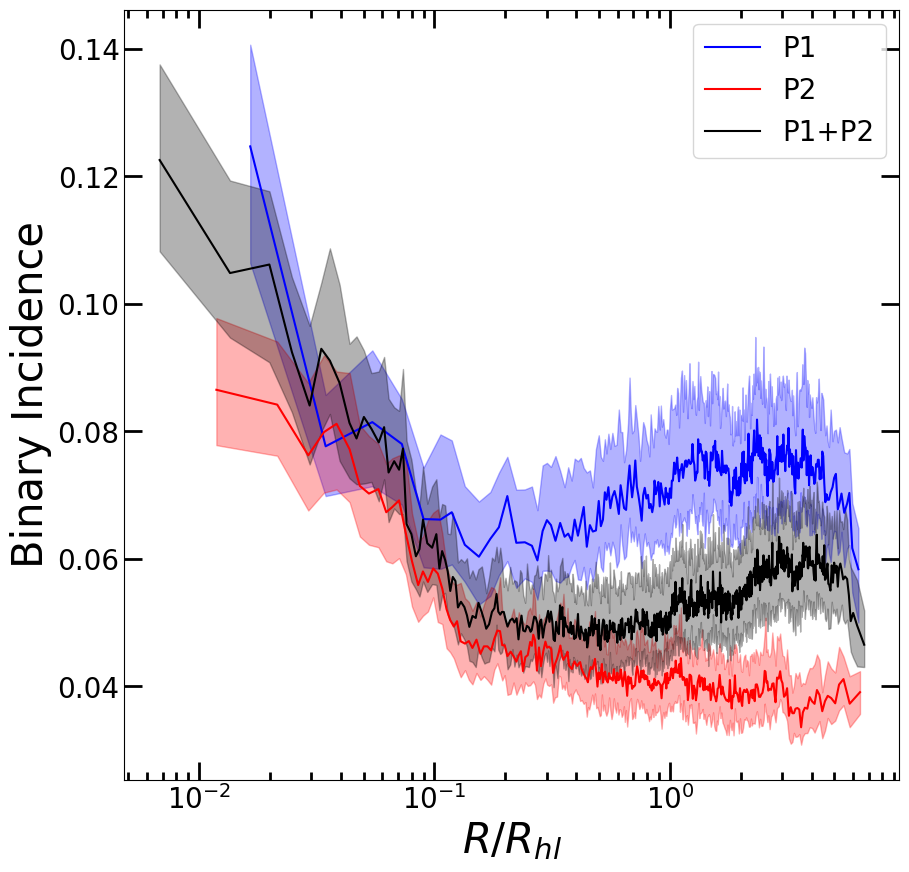}
    \caption{Radial profile of the incidence of binaries in P1 (blue), P2 (red), and both populations (black) as a function of the projected distance from the cluster center normalized to the projected half-light radius for the mr025c005fb10 simulation evolved to 12 Gyr. We report the median of 100 random realizations of the 2D spatial projection, with the shaded regions representing the 25th and 75th percentiles. Incidence refers to the fraction of objects within each population that are binaries (see Eq. 3).}
    \label{fig:fig 4}
\end{figure}

Figure \ref{fig:fig 4} shows the projected 2D radial profile of binary incidence for the simulation mr025c005fb10 at 12 Gyr. We generated 100 random realizations of the 2D projected location of objects in each simulation and report the median, presenting the 25th and 75th percentiles as the shaded regions. The P2 binary incidence declines with increasing radius as the more massive P2 binaries segregate toward the inner regions and remain centrally concentrated, while lower-mass single stars migrate outward due to the effects of two-body relaxation \citep{HeggieHut2003}. The P1 distribution shows a decrease with radius, but also a slight increase toward the very outskirts of the cluster. This can be attributed to the lower stellar density and longer relaxation timescales in the outskirts slowing migration and allowing a higher fraction of binaries to survive.

\begin{figure}
    \centering
    \includegraphics[width=0.95\linewidth]{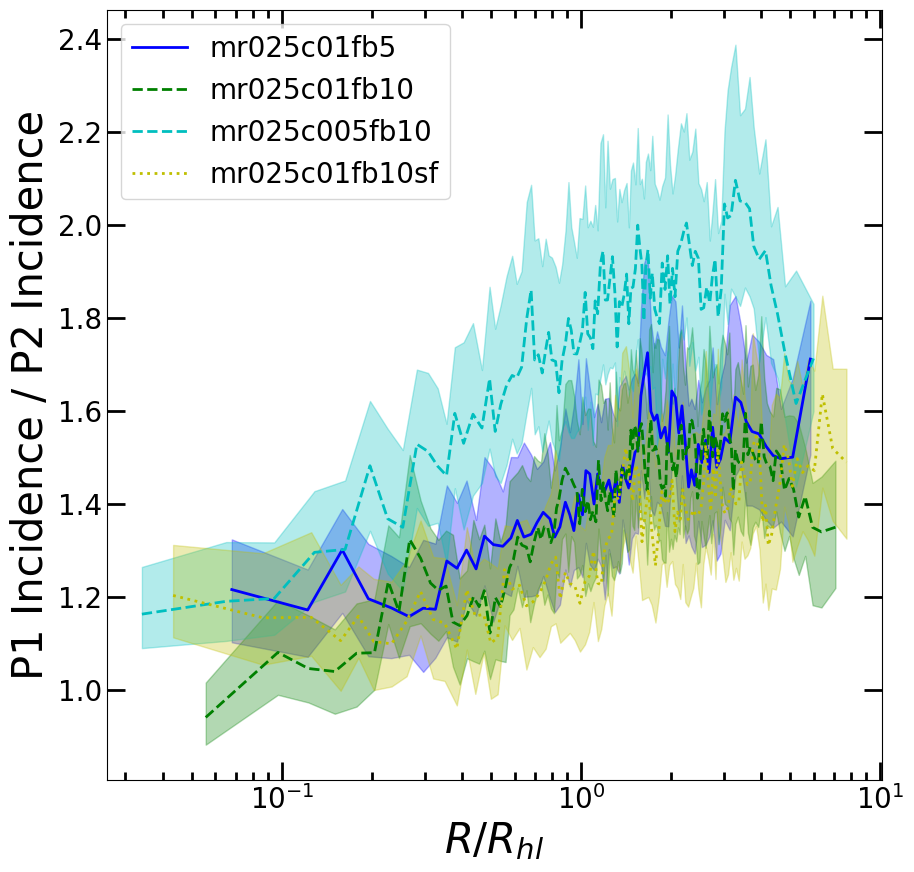}
    \caption{Radial profile of the ratio of the incidence of binaries in P1 to the incidence in P2 at 12 Gyr. We report the median of 100 random realizations of the 2D spatial projection, with the shaded regions representing the 25th and 75th percentiles. Incidence refers to the fraction of objects within each population that are binaries (see Eq. 3).}
    \label{fig:fig 5}
\end{figure}

The difference in these radial profiles is highlighted in Figure \ref{fig:fig 5}, which shows the radial profile of the binaries' incidence ratio (P1/P2). In the inner regions, where the binaries migrate toward, the ratio approaches values of $\sim1-1.2$. At greater distances from the center, the ratio continues to increase, emphasizing the difference in the radial profiles of the multiple populations. The outer regions of the cluster are therefore the most informative regions for distinguishing the incidences of the two populations. 

We also find that the simulations modeled in the stronger tidal field and undergoing stronger tidal stripping exhibit the flattest distributions. The enhanced tidal stripping preferentially removes stars from the outer regions of the cluster, exposing more centrally concentrated stars and reducing the P1/P2 incidence ratio compared to clusters at greater galactocentric distances. This is indicative of the older dynamical age of this cluster, resulting in the populations experiencing a much larger degree of spatial mixing.

\begin{figure}
    \centering
    \includegraphics[width=0.95\linewidth]{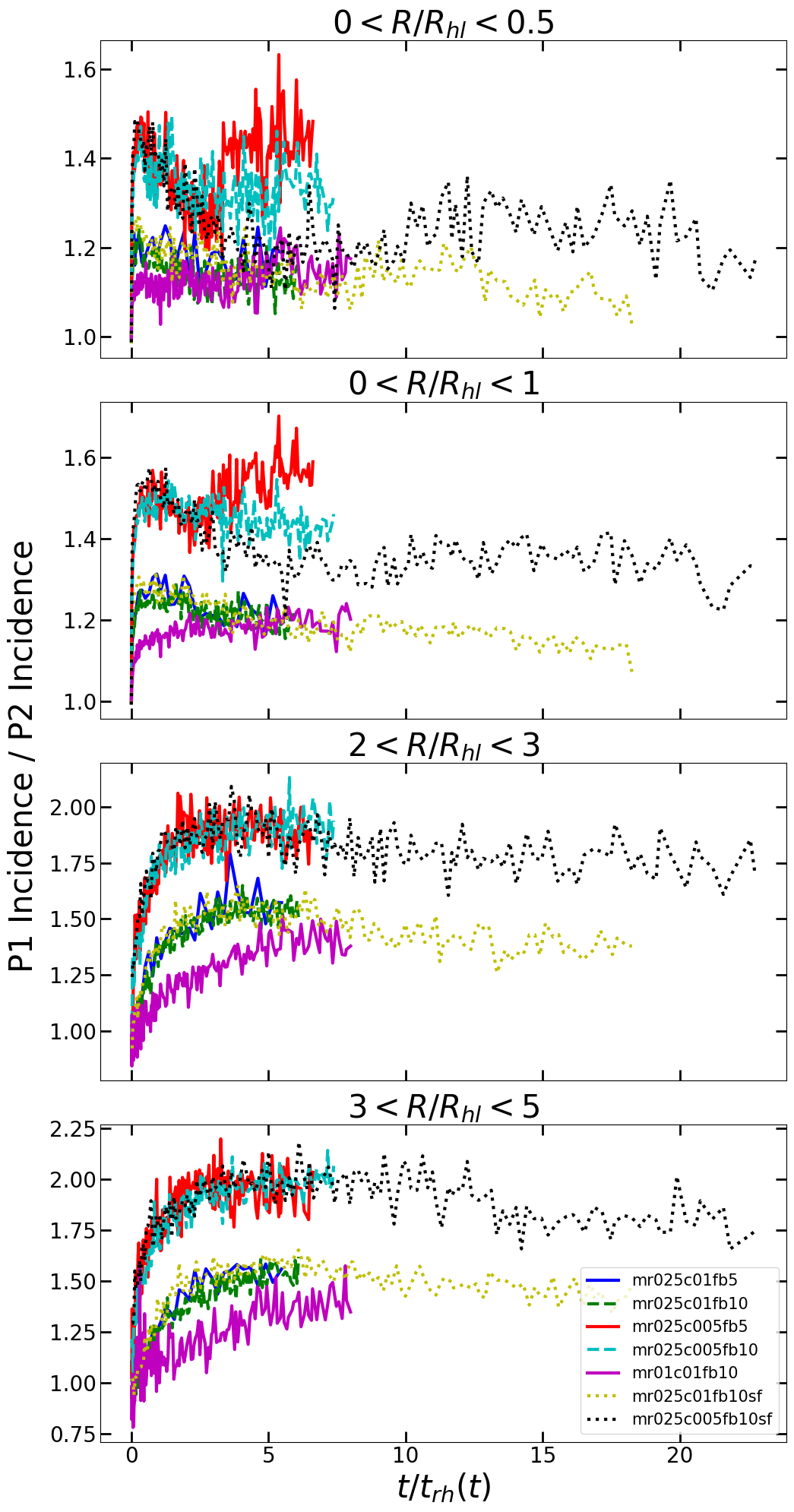}
    \caption{Time evolution of the ratio of the local P1 binary incidence to the local P2 binary incidence in four radial regions; within 0.5 projected half-light radii, within 1 projected half-light radius, between 2 and 3 projected half-light radii, and between 3 and 5 projected half-light radii, from top to bottom respectively. Incidence refers to the fraction of objects within each population that are binaries (see Eq. 3).}
    \label{fig:fig 6}
\end{figure}

Figure \ref{fig:fig 6} shows the time evolution of the ratio of the P1 binary incidence to the P2 binary incidence within four distinct radial regions. In the innermost regions, the ratio initially increases but then halts and slightly decreases to approach values equal to $\sim 1-1.2$ as the cluster evolves. In contrast, the outer regions exhibit a more pronounced and sustained increase in the value of the ratio, with P1 maintaining a higher binary incidence compared to P2 for the duration of the simulations. This increased ratio in the outer regions displays the fact that many P2 binaries migrating outward are disrupted before reaching the less hostile outer regions, as well as the preferential outward migration of single stars compared to binary stars. Additionally, P2 binaries tend to occupy more radial orbits compared to P1 binaries, increasing their disruption rate as they travel closer to the cluster center at pericenter \citep{Pavlik+2022}.

\begin{figure}
    \centering
    \includegraphics[width=0.95\linewidth]{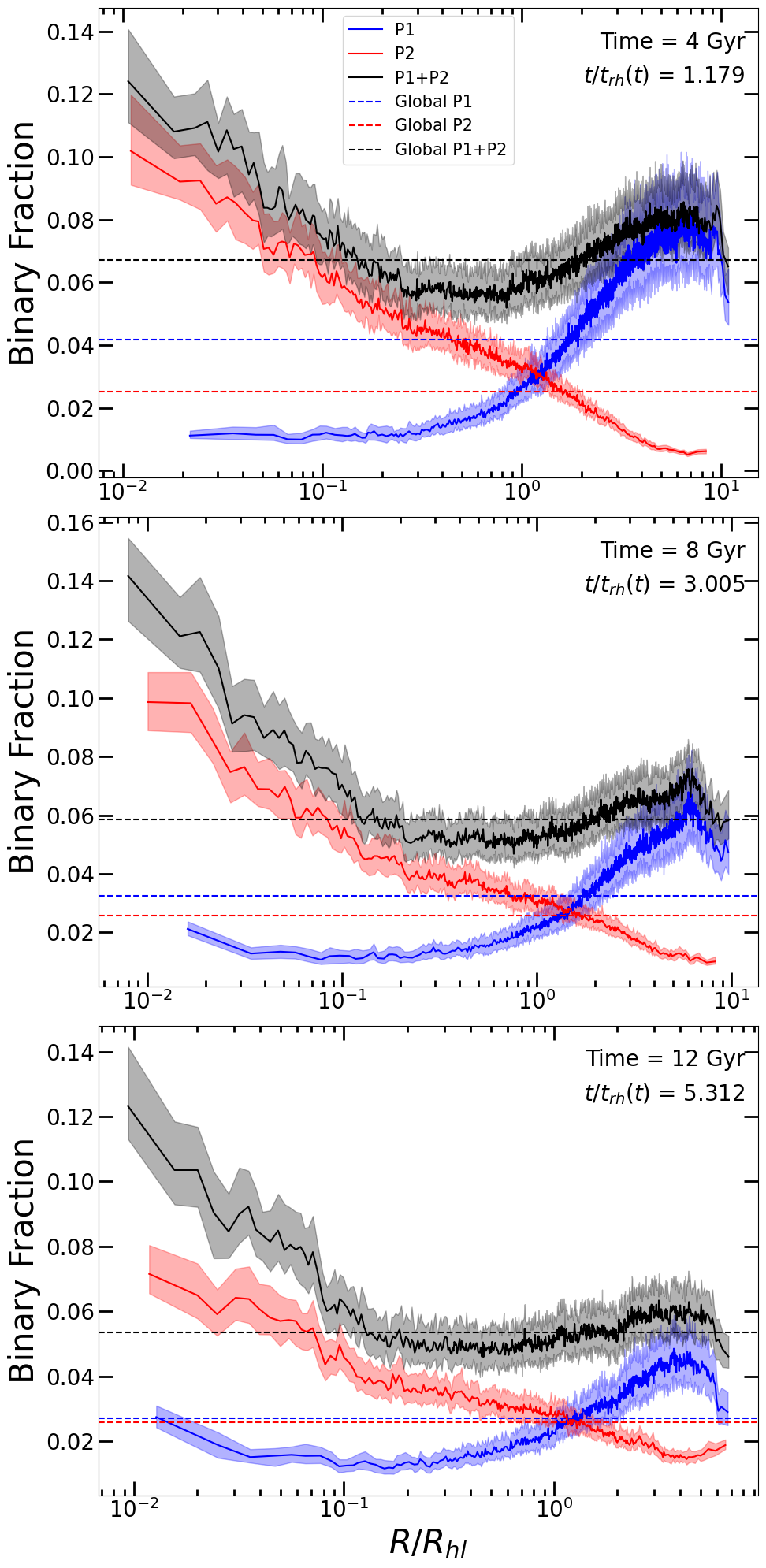}
    \caption{Radial profile of the fraction of binaries in P1 (blue), P2 (red), and both populations (black) evolved to 4 Gyr (top), 8 Gyr (middle), and 12 Gyr (bottom) for simulation mr025c005fb10. We report the median of 100 random realizations of the 2D spatial projection, with the shaded regions representing the 25th and 75th percentiles. The horizontal dashed lines represent the global value of the binary fraction for P1 (blue dashed line), P2 (red dashed line), and P1+P2 (black dashed line). Binary fraction refers to the fraction of objects within the total system that are binaries (see Eq. 4).}
    \label{fig:fig 7}
\end{figure}

We now turn our attention to the evolution of the binary fraction. The binary fraction for P1 is defined as

\begin{equation}
    f_b^{P1} = \frac{N_b^{P1}}{N_b^{Tot} + N_s^{Tot}}
\end{equation}

\noindent which is the fraction of total objects ($N_b^{Tot} + N_s^{Tot}$) in the cluster that are binaries from P1 ($N_b^{P1}$). An analogous definition once again holds for P2.

Figure \ref{fig:fig 7} shows the radial variation of the binary fraction at 4, 8, and 12 Gyr. At 4 Gyr, the binary fraction displays a profile that increases in both the cluster center and in the outskirts, with a minimum near the half-light radius (see, e.g., \citeauthor{Cordoni+2025} \citeyear{Cordoni+2025} for an observational study of this binary fraction increase in the outer region of 47 Tuc). This dual increase reflects the prevalence of the two populations in each region. Despite the higher disruption rates, P2 binaries dominate the inner binary fraction; this is in part the result of the initial inner concentration of the P2 populations and in part due to the inward migration of binaries which proceeds more rapidly for the P2 population. Concurrently, the higher survival rate of P1 binaries at larger radii, along with the longer local relaxation timescales slowing their inward migration, maintains a larger binary fraction in the outskirts. In the intermediate regions, the inward migration of central binaries alongside the insufficient replenishment of binaries from the outer regions causes a minimum to be reached. This minimum is found to varying degrees in all the simulations.

As the cluster evolves, the larger binary fraction in the outer regions begins to diminish. By 12 Gyr, the radial profile becomes significantly flatter at $ R \gtrsim 0.3\;R_{hl}$. This flattening results from the inward migration and eventual disruption of the outermost binaries, the limited outward migration of binaries from the inner regions, and the enhanced tidal stripping that preferentially removes bound systems from the cluster outskirts.

The complex dynamical evolution of the multiple populations results in a distinct and unique radial profile for the binary fraction. The presence of this distinct profile indicates that the local binary fraction can differ significantly from the global cluster value. It is interesting to note that once the binary fraction flattens in the outer regions, the value of the local (P1 + P2) binary fraction at $R>R_{hl}$ is approximately equal to the global binary fraction at the corresponding time.

\subsection{Spatial mixing and binding energy evolution}
\label{sec:3.2}

\begin{figure*}
    \centering
    \includegraphics[width=0.8\linewidth]{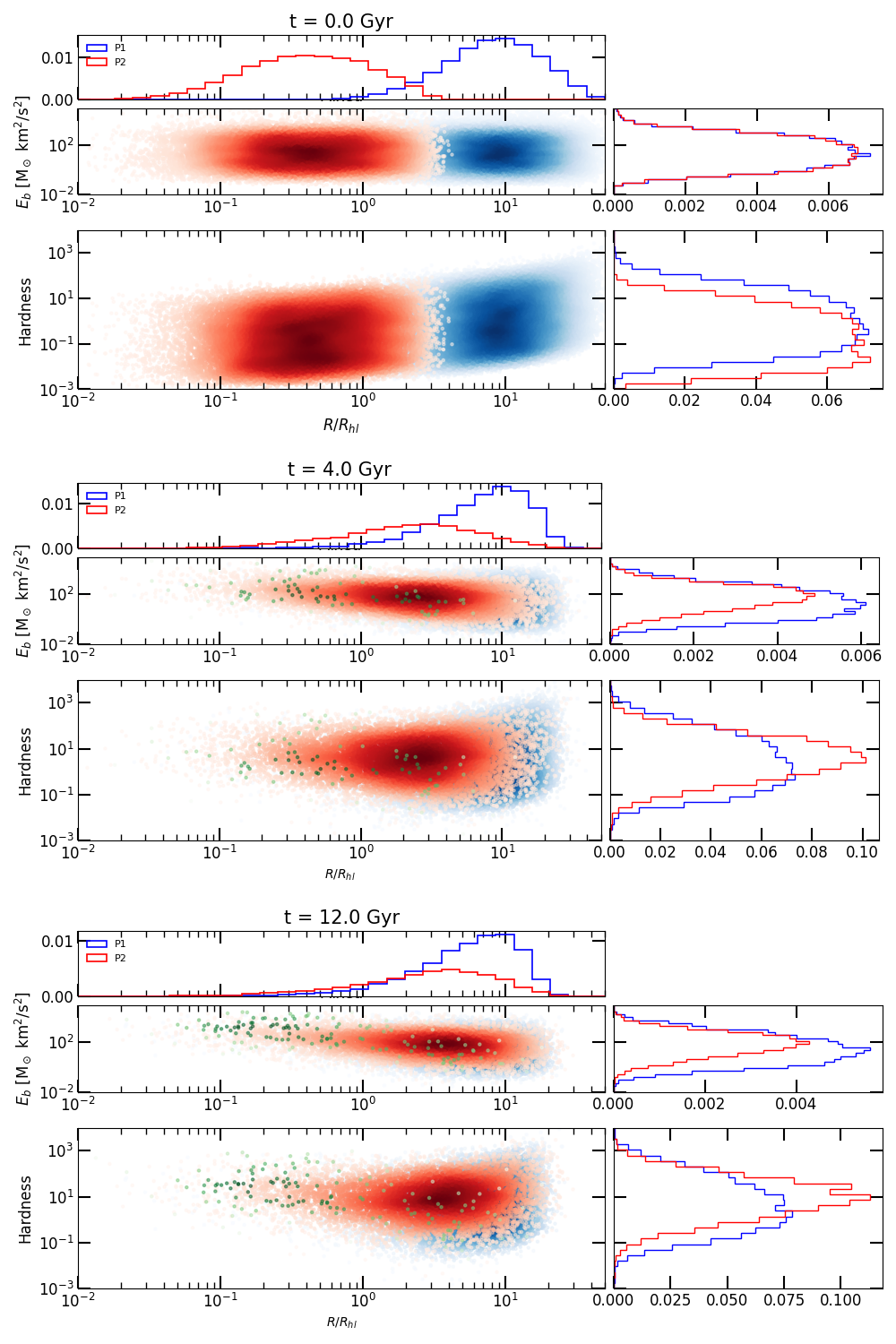}
    \caption{Radial density distributions of P1 binaries (blue), P2 binaries (red) and mixed binaries (green) compared to binding energy (top panel) and hardness (bottom panel) for simulation mr025c005fb10 at 0 Gyr (top plot), 4 Gyr (middle plot), and 12 Gyr (bottom plot). The top histogram shows the binaries in radial bins. The right side histograms show the binary distributions in binding energy bins (upper histogram) and hardness bins (lower histogram). All histograms are normalized such that the total area under the histogram equals 1.}
    \label{fig:fig 8}
\end{figure*}

In this section we discuss the evolution of the P1 and P2 binary stars on the plane of binding energy-projected clustercentric distance and investigate the spatial mixing of the multiple populations, with a focus on the difference in timescales for the mixing of binary systems relative to single stars. Figure \ref{fig:fig 8} displays the evolution of the absolute value of the binding energy and local hardness values for binaries in each population at 0 Gyr, 4 Gyr, and 12 Gyr for simulation mr025c005fb10.

At 0 Gyr (top of Figure \ref{fig:fig 8}), the two populations have binaries drawn from identical semimajor axis distributions giving comparable binding energies, but P2 is initially more centrally concentrated shifting its hardness distribution to smaller and softer values. As the cluster evolves, mass segregation and dynamical interactions drive the populations to spatially mix. In the middle panel of Figure \ref{fig:fig 8} at 4 Gyr, the initially more central P2 binaries begin migrating outward, while P1 binaries migrate inward more slowly. As discussed previously, the higher interaction rates in the dense core more efficiently disrupts the P2 binaries with lower binding energies and hardens the more compact binaries relative to P1. This results in a deficit of low binding energy binaries and an excess of hard binaries in P2, while in contrast P1 remains abundant in softer and lower binding energy binaries.

\begin{figure}
    \centering
    \includegraphics[width=0.95\linewidth]{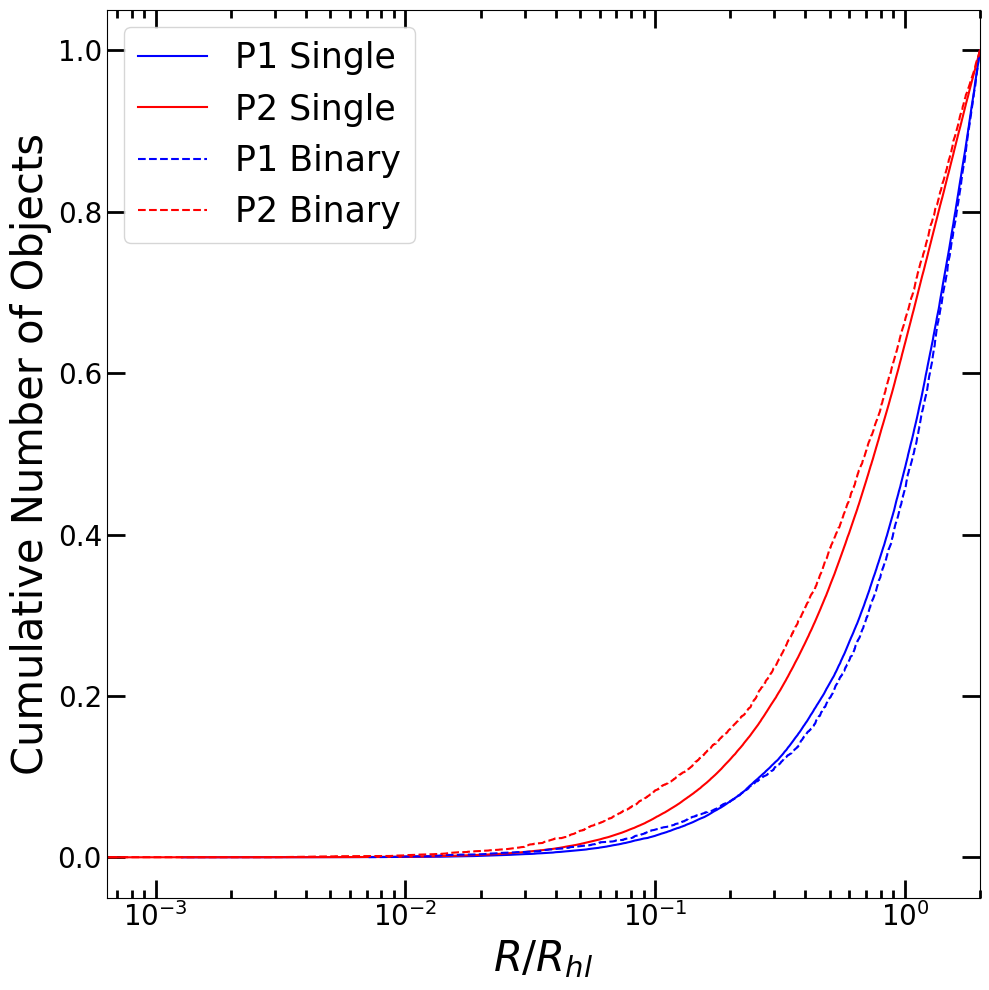}
    \caption{Cumulative radial distributions for P1 (blue) and P2 (red) for binaries (dashed line) and single stars (solid line) for simulation mr025c005fb10 at 12 Gyr.}
    \label{fig:fig 9}
\end{figure}

The dynamical evolution continues through to 12 Gyr (bottom of Figure \ref{fig:fig 8}), where the divergence in distributions becomes more apparent. P2 retains a high fraction of hard binaries while P1 preserves the initial distributions to a higher degree through the lower interaction rate. Despite the 12 Gyr of evolution and spatial mixing (see also Figure \ref{fig:fig 9} for spatial mixing of single stars), the two populations of binaries still remain discernible as P2 remains more centrally concentrated in general while P1 dominates the outer regions. Since P1 (and therefore P1 binaries) is initially less centrally concentrated than P2, many P1 binaries lie beyond several half-light radii where the velocity dispersion is lower, resulting in a larger initial fraction falling in the hard-binary regime. As these binaries migrate inward, they can transition from the hard-binary regime to the soft-binary regime as the environment harshens and the local velocity dispersion increases, enhancing their likelihood of disruption. The rate of soft binary disruption is also proportional to the local relaxation time \citep{Hut+1983}. This leads to a noticeable radial gradient showing soft binaries surviving in higher abundances at increasing radii. 

The incomplete spatial mixing of the two binary populations after 12 Gyr is further displayed in Figure \ref{fig:fig 9}, which shows the cumulative radial distributions of both binaries and single stars in each population. The single stars achieve a larger degree of spatial mixing compared to the binary stars, which more significantly retain memory of their initial distributions. The P2 binaries remain more centrally concentrated relative to P2 single stars, while the P1 binaries remain at larger radial distances compared to single P1 stars. This indicates a delayed spatial mixing timescale for binaries compared to single stars in the cluster.

To numerically quantify the degree of spatial mixing, we adopt the commonly used $A+$ parameter \citep{Alesandrini+2016, Lanzoni+2016},

\begin{equation}
    A^+ = \frac{1}{r_{\text{max}}} \int_0^{r_{\text{max}}} \left( \phi_{P1}(r') - \phi_{P2}(r') \right) \, dr'
\end{equation}

\noindent where $\phi_{P1}$ and $\phi_{P2}$ are the cumulative radial distributions for P1 and P2, respectively, normalized to an arbitrarily chosen maximum radius (2 half-light radii ($R_{hl}$) in this work). As the populations spatially mix, the $A+$ value tends to zero.

\begin{figure}
    \centering
    \includegraphics[width=0.95\linewidth]{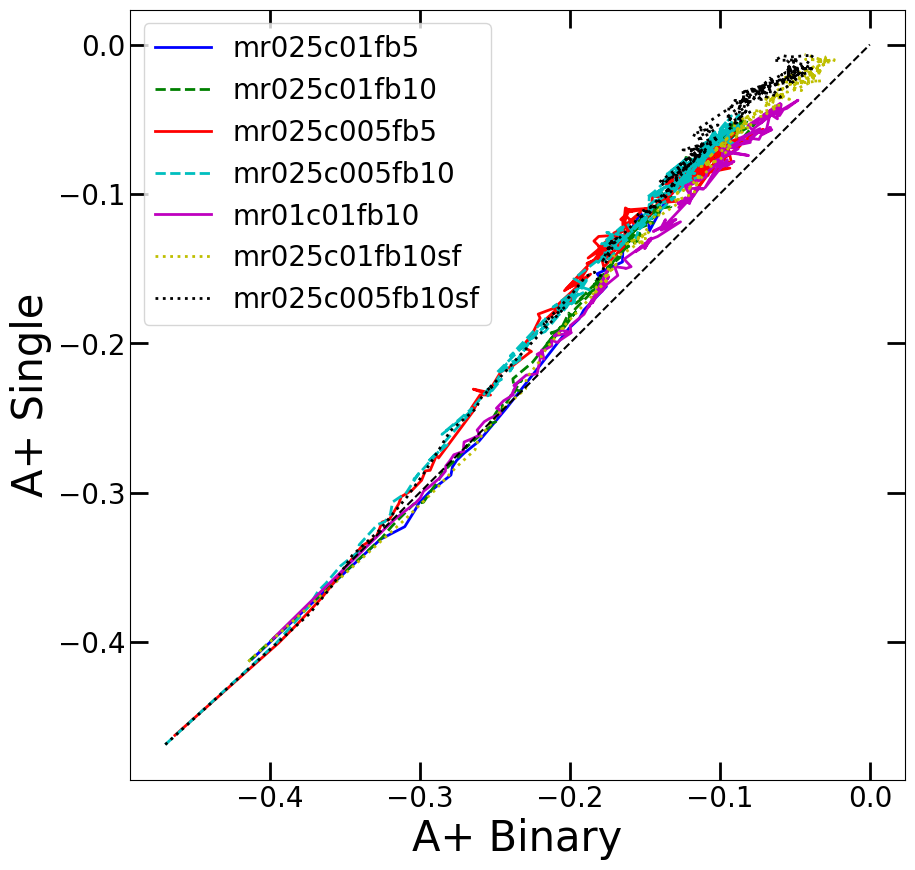}
    \caption{Evolution of the value of the $A+$ parameter for binaries (x-axis) and single stars (y-axis) in all simulations. Simulations begin in the bottom left, and evolve toward the top right of the plot over time.}
    \label{fig:fig 10}
\end{figure}

Figure \ref{fig:fig 10} shows the evolution of the $A+$ parameter for binary and single stars as the simulations evolve. The single star populations mix rapidly, reaching small $A+$ values within a few gigayears of evolution. In contrast, binary stars mix at a slower rate, retaining more negative $A+$ values at all stages of cluster evolution. Even after 12 Gyr, the binaries exhibit significantly less mixing relative to single stars. A recent observational study by \citet{Bortolan+2025} reported a similar trend in the Galactic globular cluster NGC 288, where the binary populations display a lesser degree of spatial mixing compared to single stars, consistent with the results of this paper (see also \citeauthor{Dalessandro+2018} \citeyear{Dalessandro+2018} for an observational study of NGC 6362 showing the possible kinematic effects of differences between the degree of mixing in P1 and P2 binary and single stars).

\begin{figure}
    \centering
    \includegraphics[width=0.95\linewidth]{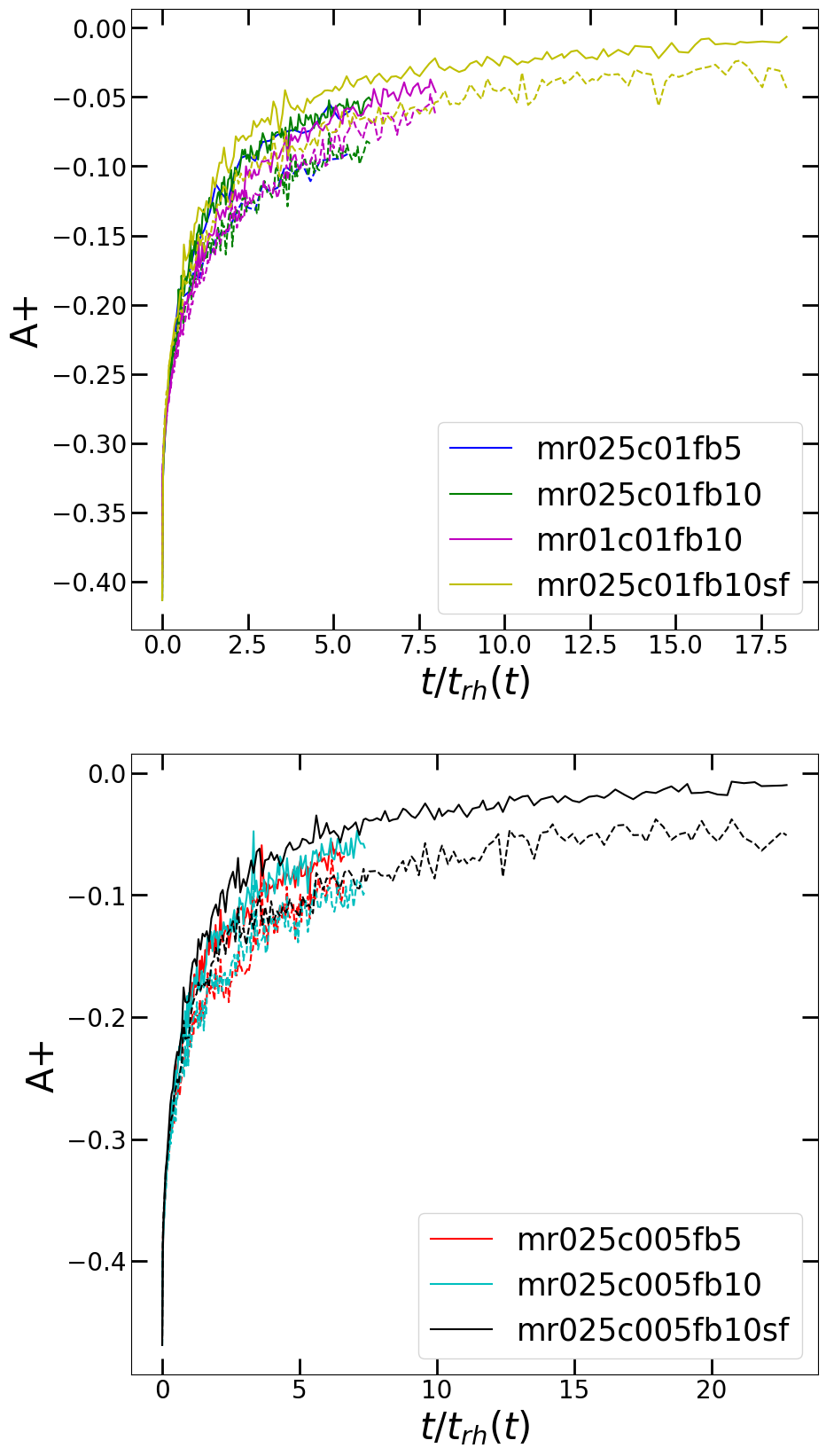}
    \caption{Time evolution of the $A+$ parameter for single stars (solid lines) and binary stars (dashed lines) in the simulations with a higher concentration parameter (top panel) and a lower concentration parameter (bottom panel).}
    \label{fig:fig 11}
\end{figure}

Figure \ref{fig:fig 11} shows the time evolution of the $A+$ value for binary and single star populations in all simulations. In every case, the binary populations display more negative $A+$ values than single stars, pointing toward slower binary spatial mixing in general. Additionally, in the simulations mr025c01fb10sf and mr025c005fb10sf, which attain much higher dynamical ages, the binary stars still show a difference in spatial distribution when the single stars have attained nearly complete mixing. This dynamical feature is caused by several mechanisms that are unique to stellar multiples, such as binary ionization, ejection and recoil \citep{Hong+2015, Hong+2016}. Additionally, P2 binaries segregate inward more efficiently than P2 single stars in general due to their higher combined mass, although only the most compact binaries will survive the dense center. Ionization and ejection events act to balance the inward migration of P1 binaries by destroying some systems and displacing others outward, maintaining a lower P1 binary fraction in the inner regions of the cluster and thereby reducing the degree of spatial mixing. In addition, P1 binaries are initially located in the outer regions of the cluster where the relaxation time is longer, causing their inward segregation to proceed more slowly than P2 binaries even for systems of comparable mass. The combination of these dynamical processes result in the longer spatial mixing timescale for binary stars, preserving spatial fingerprints of the initial population radial distributions even as the cluster evolves for long periods of time (see, e.g., \citeauthor{Hong+2019} \citeyear{Hong+2019}).

\subsection{Mixed binaries}
\label{sec:3.3}

In this section we focus on the formation and evolution of mixed binaries, which contain one stellar component from each population (P1 and P2). These binaries are not primordial, and are instead produced dynamically mainly through exchange interactions. This formation mechanism is more relevant in the inner cluster regions where stellar densities and encounter rates are highest.

We now return to Figure \ref{fig:fig 8}, which displays the evolution of the absolute value of the binding energy and local hardness values for binaries within each population at three different times for simulation mr025c005fb10. This plot shows that the mixed binaries are almost exclusively located in the innermost regions, and rarely migrate outward as the cluster evolves. Their central formation reflects their dynamical origin, and their confinement to the inner regions suggests mass segregation related effects. During an exchange interaction, the more massive component of the binary is typically retained, while the lower-mass component is ejected (see, e.g., \citeauthor{Hills1992} \citeyear{Hills1992}; \citeauthor{Heggie+1996} \citeyear{Heggie+1996}; \citeauthor{HeggieHut2003} \citeyear{HeggieHut2003}). This process generally results in a mixed binary with a larger combined mass, which is more likely to remain centrally concentrated.

An interesting dynamical feature of mixed binaries is their intermediate hardness values at the time of their formation, as seen in Figure \ref{fig:fig 8}. The intermediate values are a result of the regime limits of an exchange interaction. Binaries that are the most compact (i.e., hardest) are less likely to undergo a close interaction leading to a component exchange. In contrast, binaries with the larger semimajor axes (i.e., softer) are more likely to be disrupted during a stellar encounter or to reach the escape velocity after strong interactions, rather than having a component exchanged. As a result, the probability for forming mixed binaries peaks at intermediate hardness values \citep{Hong+2015}.

\begin{figure}
    \centering
    \includegraphics[width=0.95\linewidth]{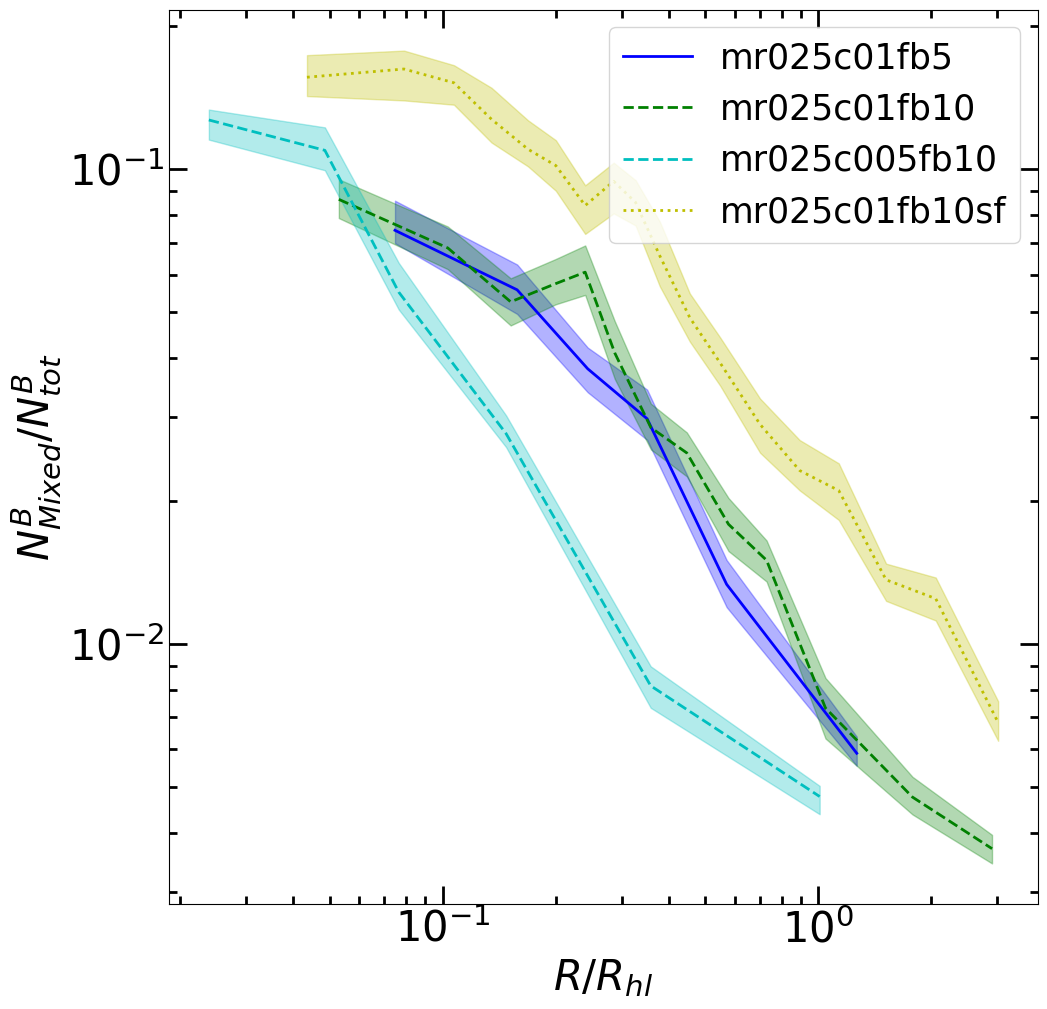}
    \caption{Radial profile of the fraction of binaries in a mixed binary configuration for simulations at 12 Gyr. We report the median of 100 random realizations of the 2D spatial projection, with the shaded regions representing the 25th and 75th percentiles.}
    \label{fig:fig 12}
\end{figure}

Figure \ref{fig:fig 12} shows the radial profile of the mixed binary fraction. This fraction displays a steep decline with increasing distance from the cluster center; most of the mixed binaries are located within one projected half-light radius. The central concentration and lack of mixed binaries in the outer regions in our simulations is consistent with the findings of the first observational studies of multiple-population binaries. \citet{Milone+2025} reported a higher fraction of mixed binaries in the central regions of 47 Tuc, compared to the lower fraction in the outer regions. Similarly, \citet{Milone+2020} find a population of mixed binaries in the central regions of NGC 288, while \citet{Bortolan+2025} observe an absence of mixed binaries in the outskirts. These findings are consistent with the results found in this paper, where mixed binaries form and remain in the central regions as globular clusters evolve.

\subsection{Main sequence--white dwarf binaries}
\label{sec:3.4}

This section presents the results from an interesting subset of the binary population: objects composed of a main-sequence (MS) star and a white dwarf (WD) star. These MS-WD binaries provide valuable insights into the complex interplay between stellar evolution and dynamical processes in multiple-population globular clusters.

\begin{figure}
    \centering
    \includegraphics[width=0.95\linewidth]{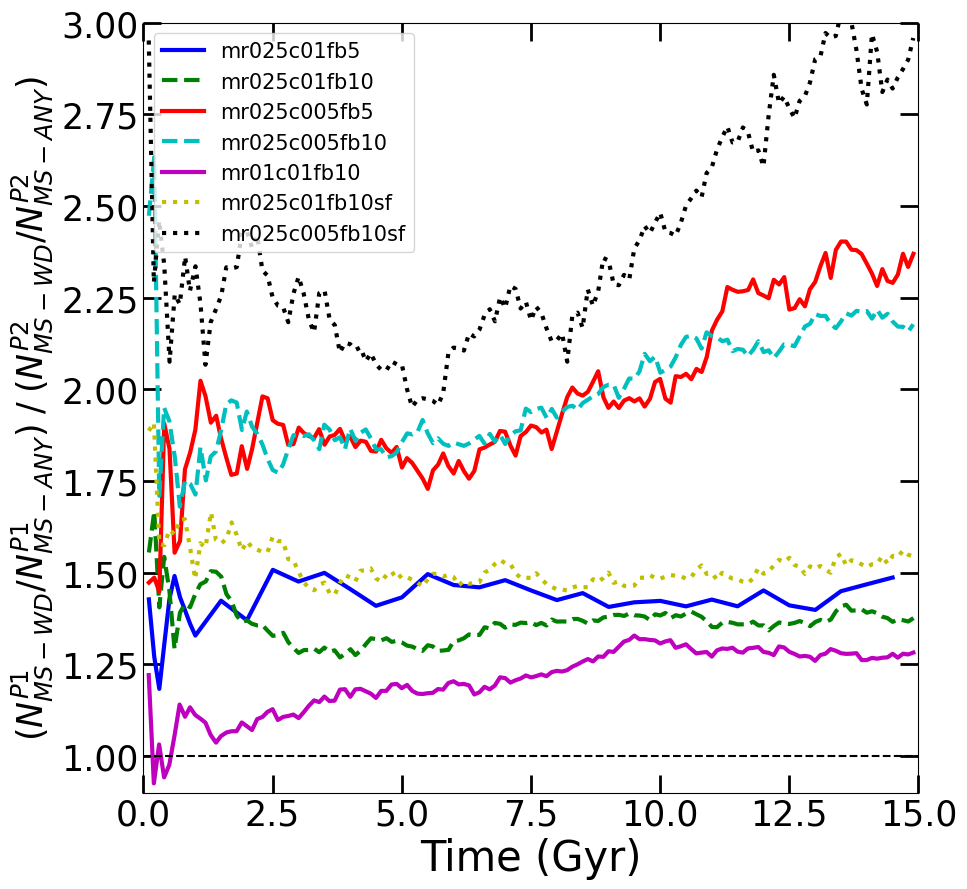}
    \caption{Time evolution of the ratio of the incidence of MS P1 binaries ($N^{P1}_{MS-ANY}$) in a MS-WD configuration ($N^{P1}_{MS-WD}$) to the incidence of MS P2 binaries ($N^{P2}_{MS-ANY}$) in a MS-WD configuration ($N^{P2}_{MS-WD}$).}
    \label{fig:fig 13}
\end{figure}

Figure \ref{fig:fig 13} shows the time evolution of the ratio of the MS-WD binary incidence in P1 to that in P2. By MS-WD binary incidence we are referring to the fraction of binaries containing at least one MS star (MS-ANY) that are in a MS-WD configuration. We define a MS-WD binary within a given population as a binary system containing a MS star from that population and a WD companion from either population. Initially, this ratio is undefined as no stars have yet evolved to become WDs. As stellar evolution proceeds and the more massive MS stars evolve off the main sequence, MS-WD binaries are formed in both populations, however, all simulations show an excess in the incidence of MS-WD binaries found in P1 compared to P2.

This increased incidence of MS-WD in the P1 binary population is likely a relic of the post-evolution hardness of the MS-WD binary. When a component of an MS-MS binary evolves into a WD, the associated mass loss can significantly alter the binary orbit, in some cases increasing the separation and decreasing the binding energy as a result (see, e.g., \citeauthor{Taurisbook2023} \citeyear{Taurisbook2023}). This can soften the binary system, and the preferential disruption of softer binaries in P2 will then result in more MS-WD binaries being disrupted compared to P1. As a result, wider MS-WD binaries are more likely to survive in the less dense outer regions occupied by P1, leading to the trend highlighted in Figure \ref{fig:fig 13}. 

Additionally, some MS-WD binaries may form through common envelope evolution, which tightens the resulting system. P1 binaries that undergo this common envelope phase will become harder and are more likely to survive the inward migration toward more hostile environments. In contrast, the progenitors of P2 MS-WD binaries are initially in the more crowded cluster center, and are more likely to be disrupted before they can undergo the common envelope phase and harden (similar to the preferential disruption of cataclysmic variable progenitors in the core of GCs investigated in \citeauthor{Davies1997} \citeyear{Davies1997} and \citeauthor{Belloni+2019} \citeyear{Belloni+2019}), further contributing to the lower incidence of MS-WD binaries in P2 compared to P1. The simulations show the ratio increasing initially, indicating the preferential P2 disruption, as well as a flattening as the populations mix and the dynamical environments of binaries in both populations become similar.

\begin{figure}
    \centering
    \includegraphics[width=0.95\linewidth]{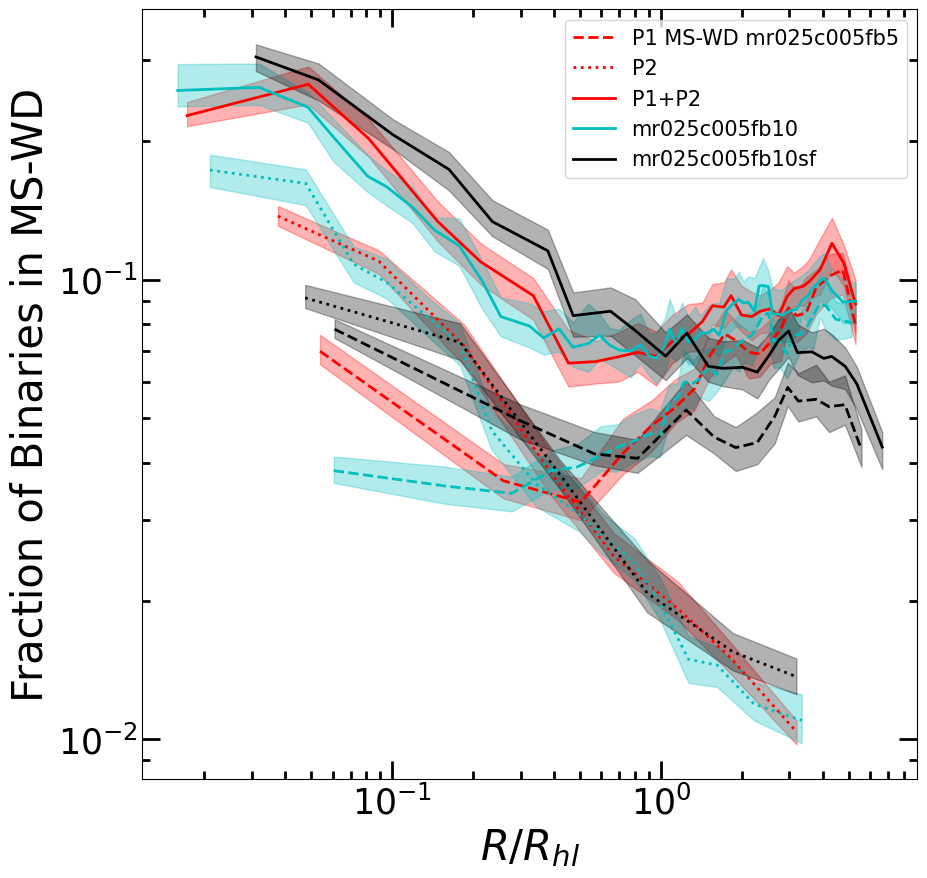}
    \caption{Radial profile of the fraction of binaries in a MS-WD configuration in P1 (dashed line), P2 (dotted line), and both populations (solid line) for the simulations at 12 Gyr. We report the median of 100 random realizations of the 2D spatial projection, with the shaded regions representing the 25th and 75th percentiles.}
    \label{fig:fig 14}
\end{figure}

In Figure \ref{fig:fig 14} we display the radial distribution of the fraction of all binaries that are composed of a main-sequence star and a white dwarf within each population and the cluster as a whole. This profile resembles the mixed binary radial profile discussed in Sect. 3.3 (Figure \ref{fig:fig 12}), with the fraction peaking in the central regions and decreasing with increasing radius. This central concentration can be explained through mass segregation, where the more massive stars migrate more efficiently and evolve to become WD at a faster rate. Additionally, the hardening of binaries post-common envelope evolution increases their likelihood of survival as they migrate inward relative to their softer MS-MS progenitors. These two processes result in binaries containing a WD being found largely in the innermost regions, and rarely in the outer regions containing the lower-mass stars that have yet to evolve off the main sequence.

The radial distribution of MS-WD binaries within P2 similarly shows a decrease with radius, but with a larger slope. This is likely a result of the shorter local relaxation timescale in the inner regions affecting the P2 binaries, which accelerates the process described above. However, the P1 binary population shows a relatively flat distribution, with a slight increase toward the outer regions. Similar to the larger fraction of P1 binaries surviving in the outer regions, the MS-WD binaries that are softened after the WD evolution are disrupted at increasing rates toward the cluster center, decreasing the fraction. Additionally the local relaxation timescales for the outer regions are much longer than in the core, preventing a significant amount of migration from occurring which would evolve the distribution to resemble that seen in P2.

\begin{figure}
    \centering
    \includegraphics[width=0.95\linewidth]{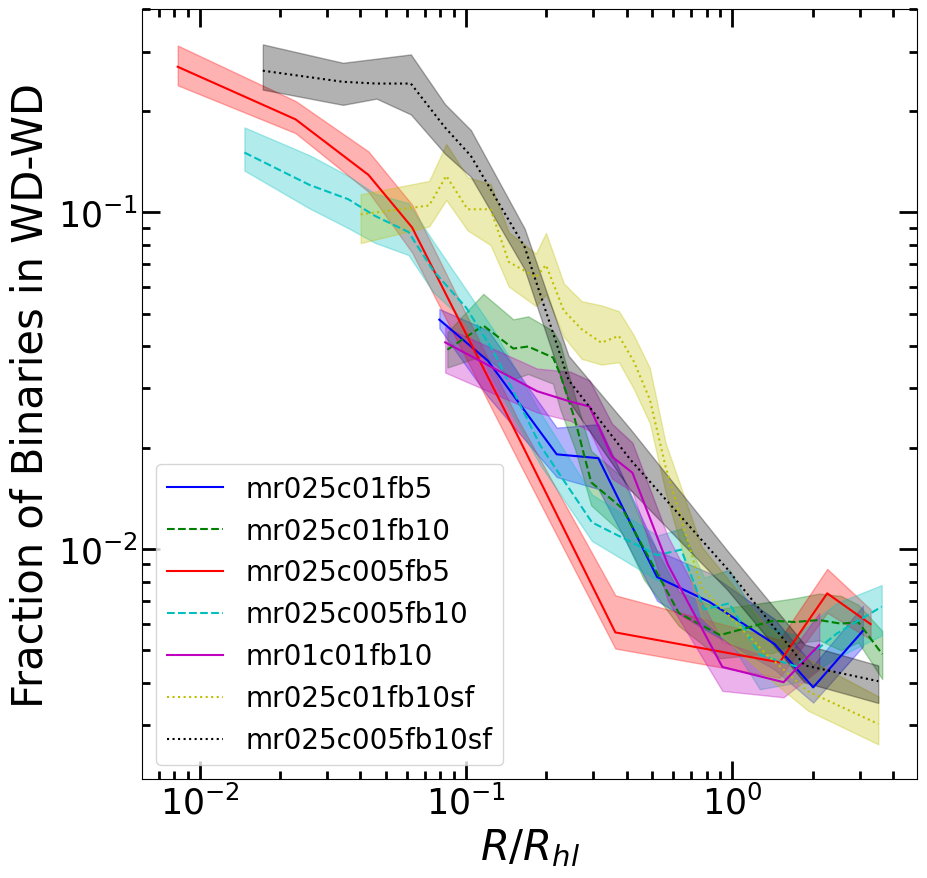}
    \caption{Radial profile of the fraction of binaries in a WD-WD configuration for the simulations at 12 Gyr. We report the median of 100 random realizations of the 2D spatial projection, with the shaded regions representing the 25th and 75th percentiles.}
    \label{fig:fig 15}
\end{figure}

In Figure \ref{fig:fig 15} we show the radial distribution of the fraction of binaries in a WD-WD binary. In all simulations there is a clear decreasing trend with radius, highlighting the centrally concentrated nature of these systems. This radial profile similarly can be explained by mass segregation, as WD-WD binaries as well as their MS progenitors are among the higher-mass objects in the system. This higher mass leads to more efficient inward migration relative to the majority of binaries, resulting in the radial profile seen. Additionally, the monotonic decrease can also be attributed in part to the formation of WD-WD binaries dynamically, similar to the formation of mixed binaries, which occurs primarily in the central regions.

\section{Conclusions}
\label{sec:4}

In this paper we presented the results from a series of Monte Carlo simulations designed to investigate the dynamical evolution of binary stars in multiple stellar populations within globular clusters. We examined the evolution of global and local binary incidences and fractions, the spatial mixing timescales of binary stars compared to single stars, and the formation and dynamical behavior of both mixed binaries and main sequence--white dwarf binaries.

Multiple theoretical and observational studies suggest that the second population (P2) of stars form more centrally concentrated than the first population (P1) of stars in globular clusters. This initial structural difference creates distinct dynamical environments for each population (P1 and P2), significantly influencing binary evolution. As a result, binary stars offer a unique probe into the early conditions and long-term dynamical evolution of the multiple populations.

The initial central concentration of P2 leads to more frequent dynamical encounters, increasing the rate of disruption and hardening for the binaries. Consequently, P2 retains a higher fraction of compact and hard binaries, whereas P1 also preserves wider systems (see Figure \ref{fig:fig 1}). This disparity produces a global excess of binaries in P1 compared to P2 (Figures \ref{fig:fig 2} and \ref{fig:fig 3}). However, the local binary incidence exhibits strong radial variation (Figures \ref{fig:fig 4} and \ref{fig:fig 5}), where the inner regions show similar incidence values between the populations, but the outer regions clearly display an increased incidence of P1 binaries. While the innermost regions evolve to have a P1/P2 incidence ratio equal to $\sim1-1.2$, the outer regions maintain a higher P1 binary incidence compared to P2 (Figure \ref{fig:fig 6}).

During the early evolutionary phases, the radial profile of the binary fraction develops a slightly bimodal shape with both the inner and outer regions showing an increase; this bimodal shape becomes less pronounced and evolves into a monotonic profile with the binary fraction decreasing with the distance from the cluster's center (Figure \ref{fig:fig 7}) as the cluster continues its dynamical evolution.

We find that the two populations of binary stars spatially mix more slowly than single stars (Figures \ref{fig:fig 9}, \ref{fig:fig 10}, and \ref{fig:fig 11}). This delayed spatial mixing can be attributed to the unique dynamical processes affecting stellar multiples such as ionization, recoil, and ejection. These results agree with observational findings and demonstrate that binaries retain spatial memories of their initial distributions longer than single stars (Figure \ref{fig:fig 8}).

Mixed binaries containing one stellar component from each population form dynamically mostly through exchange interactions, predominantly in the cluster core where encounter rates are the highest. These mixed binaries remain centrally concentrated and initially exhibit intermediate hardness values that increase with time (Figures \ref{fig:fig 8} and \ref{fig:fig 12}). The steep radial decline in number is consistent with observed trends in clusters such as 47 Tuc \citep{Milone+2025} and NGC 288 \citep{Milone+2020, Bortolan+2025}.

Finally, we explored the evolution of MS-WD binaries. The incidence of these systems is higher in P1 compared to P2 (Figure \ref{fig:fig 13}). This is due to the softening effects of post-MS mass loss and increased disruption rates in P2's denser environments, as well as the enhanced disruption of P2 MS-WD progenitors before a common envelope phase. The P2 MS-WD fraction decreases sharply with radius, whereas the P1 profile is significantly flatter and slightly increases at larger radii (Figure \ref{fig:fig 14}). This reflects the differential survival and migration of the two populations. WD-WD binaries exhibit an even more pronounced radial dependence, peaking in the central regions and decreasing sharply with radius (Figure \ref{fig:fig 15}).

Our results emphasize the importance of understanding binary stars in multiple populations, which can offer valuable insights into the initial conditions and subsequent evolution of globular clusters.

\begin{acknowledgements}

EV acknowledges support from NSF grant AST-2009193. This research was supported in part by Lilly Endowment, Inc., through its support for the Indiana University Pervasive Technology Institute. AA acknowledges support for this paper from project No. 2021/43/P/ST9/03167 co-funded by the Polish National Science Center (NCN) and the European Union Framework Programme for Research and Innovation Horizon 2020 under the Marie Skłodowska-Curie grant agreement No. 945339. This research was funded in part by NCN grant number 2024/55/D/ST9/02585. For the purpose of Open Access, the authors have applied for a CC-BY public copyright license to any author Accepted Manuscript (AAM) version arising from this submission. AH was supported by "Excellence Initiative – Research University", POB4 - competition Nr: 139 from 2024-03-28 (139/06/POB4/0030). MG was supported by the Polish National Science Center (NCN) through the grant 2021/41/B/ST9/01191.

\end{acknowledgements}

\bibliographystyle{aa}
\bibliography{ref}

\end{document}